# Beyond Prime Farmland: Solar Siting Tradeoffs for Cost-Effective Decarbonization

Papa Yaw Owusu-Obeng [a], Mai Shi [a,*], Max Vanatta [a], Michael T. Craig [a, b]

[a] School for Environment and Sustainability, University of Michigan, Ann Arbor, MI 48109, USA

[b] Department of Industrial and Operations Engineering, University of Michigan, Ann Arbor, MI 48109, USA

*Corresponding author: Mai Shi, maish@umich.edu

## ABSTRACT

The feasibility and cost-effectiveness of continued growth in solar photovoltaics are closely tied to siting decisions. But trade-offs between costs and technical potential between land categories, especially brownfields and rooftop sites, have not been quantified, despite increasing resistance to and policy interest in reducing use of greenfield sites (e.g., prime agricultural lands). We examine the effect of siting decisions across land types for utility-scale and rooftop PV on the feasibility and cost of meeting solar deployment targets across the Eastern U.S. We build a database of solar PV supply curves by land type for each county in the Eastern Interconnect (EI) region (~2,400 counties). Our supply curves quantify technical potential versus levelized cost across greenfield, brownfield, and rooftop land types. With these supply curves and a 2035 solar deployment target (435 GW) aligned with a decarbonized power system, we quantify cost and capacity trade-offs using scenarios that prioritize solar PV deployment on different land types. We find greenfield, particularly prime agriculture, sites offer the lowest levelized costs for meeting capacity targets, of 39 to 57 $/MWh. Contaminated lands, often prioritized in policy to reduce land use conflict, have limited technical potential and impose a cost premium of 14-33% relative to greenfield sites. Rooftop PV provides enough technical potential for meeting capacity targets but comes at consistently higher costs, with minimum LCOEs of roughly 70 $/MWh or well above the highest-cost greenfield sites. Our results detail heterogeneous siting trade-offs across the Eastern United States, enabling targeted policy design to meet deployment targets while balancing land use conflicts.

**INTRODUCTION**

Over the past decade, the United States has witnessed a rapid expansion in solar photovoltaic (PV) capacity, driven by technological advancements, declining costs, and enabling policies [1], [2]. In 2023, the United States added approximately 18.5 GW of utility-scale solar capacity—a 77% increase from the previous year—surpassing wind and other renewable energy technologies in new capacity additions. Looking ahead, more than 1,000 GW of proposed projects dominate the interconnection queues, indicating that large-scale solar will continue to play a pivotal role in the ongoing transition toward decarbonized power systems[2]. However, as deployment accelerates, questions surrounding how land-use choices influence land availability, solar technical potential, and the costs of solar PV deployment have become increasingly critical[3], [4], [5], [6].

The feasibility and cost-effectiveness of utility-scale PV deployment are closely tied to land-use decisions[6], [7]. Certain land categories, such as contaminated or disturbed lands, offer advantages by minimizing land-use conflicts and environmental permitting hurdles[8], [9]. However, these sites may also introduce higher infrastructure costs and specialized engineering requirements[10], [11]. Similar tradeoffs emerge for other land categories. Agricultural lands, while abundant, may require balancing solar generation with food supply considerations and local community interests[12], [13], [14]. Barren and non-agricultural lands may reduce competing land-use pressures but often face challenges related to grid access, which increases interconnection costs[4], [15]. Even rooftop spaces in urban settings, though they ease demands on greenfield sites, pose their own engineering and cost hurdles[16], [17], [18]. Balancing these tradeoffs to meet state and regional decarbonization goals requires an understanding of the technical potential, the overall expense across these diverse land categories, and policy considerations associated with different land categories.

Existing studies have examined how land-use considerations shape regional solar potential, employing geospatial analyses that factor in slope, zoning regulations, environmental protections, and other exclusion criteria [19], [20], [21], [22], [23], [24]. For example, stringent siting restrictions, such as prohibiting development on prime farmland and intact grasslands, have been shown to reduce solar capacity potential by 40–62% in the Western U.S. [19], [20]. Similarly, regulatory exclusions from federal infrastructure lands have been estimated to limit solar potential by up to 53% [24]. At the national level, zoning regulations alone can reduce available solar-suitable land by 38% across the U.S. [23] and by as much as 54% in the Great Lakes region [22]. Beyond technical potential, land-use restrictions have been found to have significant implications for the cost of solar deployment. By limiting site selection, restrictions may force developers to locate projects in areas with lower solar resource quality, increasing overall system costs. For instance, Wu et al. [20] found that prohibiting solar development in environmentally sensitive habitats led to a 3% increase in system-wide costs due to reduced access to high-insolation sites. Similarly, Mai et al. [24] showed that limiting solar deployment on federal lands can exhaust the most cost-effective sites early, driving up the costs of remaining developable locations.

While these studies highlight the general impact of siting restrictions on solar capacity and costs, detailed breakdowns of cost trade-offs and technical potential associated with individual land categories, especially brownfields, barren and rooftop sites, have remained unexplored. Moreover, no research to date has systematically examined how prioritizing one land-use type over another, or combining multiple land-use categories, might impact solar deployment costs and capacity mandates in meeting regional decarbonization targets.

In this study, we examine the effect of siting decisions across land types for utility-scale and rooftop PV on the feasibility and cost of meeting solar deployment targets across the Eastern U.S. We specifically distinguish among building sizes (large, medium, small rooftops), greenfield sites (agricultural, forests, shrubland, range/grasslands, and barren), and contaminated sites (superfund sites, abandoned mine lands, landfills, brownfields and RCRA Corrective Action sites). We first create a comprehensive geospatial database by collecting and characterizing solar land-use and siting constraints across 2,400 county-level jurisdictions covering the Eastern U.S. This database captures an unprecedented level of detail on where solar facilities can be developed within each land category, surpassing the geographic breadth and depth of previous analyses. We then couple these geospatial data with cost and performance assumptions to generate land-specific supply curves of potential solar development. To evaluate the impact of land-use choices on future solar deployment, we analyze scenarios that vary the priority assigned to different land categories for solar development. By comparing model outcomes across these land-category scenarios, we provide critical insights into the tradeoffs between land allocation, solar resource potential, and the overall cost of decarbonizing the power sector.

# METHODS

We begin by quantifying county-level land availability for utility-scale solar PV across different greenfield and contaminated land categories, and rooftop availability for distributed solar PV across rooftop sizes. We then translate land availability to capacity and calculate the levelized cost of electricity (LCOE) for solar PV across counties and land types (which refers to land use types for utility-scale PV and rooftop sizes for distributed PV). Aggregating costs and capacities across counties and land types to the Eastern Interconnection and constituent regional transmission organization (RTO) footprints yields land-type-specific supply curves. We combine these supply curves with future solar PV deployment targets in a scenario analysis that quantifies cost trade-offs for PV deployment under different land type prioritizations.

## Spatial characterization of land available for utility-scale solar across land types

To estimate the availability of land for utility-scale solar deployment, we develop a geospatial database of land categories and apply site suitability constraints across roughly 2,400 counties in the Eastern Interconnection using ArcGIS. First, we identify potential greenfield lands from the 2020 U.S. Geological Survey's National Land Cover Dataset (NLCD), which provides a 30m resolution of land cover types derived from satellite imagery [25]. Greenfield land categories in the dataset include cropland, forest, shrubland, rangeland/grassland, and barren/marginal lands. Because the NLCD does not explicitly classify contaminated or previously disturbed sites, we obtain data on brownfield sites from the U.S. Environmental Protection Agency (EPA)'s RE-Powering America's Land Initiative [26]. See Table S1 in the SI for the description of land types. This dataset includes the georeferenced location and estimated developable acreage of contaminated lands, including Superfund sites, designated under the Comprehensive Environmental Response, Compensation, and Liability Act (CERCLA) for cleanup of hazardous contamination; abandoned mine lands; landfills; Resource Conservation and Recovery Act (RCRA) Corrective Action sites, which are locations undergoing or having completed remediation of hazardous waste facilities; and brownfields. Both Superfund and RCRA Corrective Action sites are increasingly being repurposed for renewable energy development [27]. To ensure accurate classification, we cross-reference contaminated sites in the RE-Powering dataset with overlapping

NLCD categories. Sites listed as contaminated but located within greenfield land types (e.g., barren land or grassland) in the NLCD are removed from the greenfield pool to avoid double-counting.

Next, we apply a uniform set of exclusion criteria to each land cover type to determine site suitability for solar. This removes physical constraints such as slope thresholds (>10° for solar) [19], [28], and protected areas including wetlands, conservation easements, and designated ecological zones. We also exclude areas within buffer distances of airports, military bases, and urban areas to reflect typical permitting restrictions. Finally, we group developable land for solar by county for each land type category. The final dataset allows a spatial characterization of where utility-scale solar projects are feasible under various land-use scenarios.

## Spatial characterization of rooftop area available for solar across building sizes

We estimate available rooftop area across the Eastern United States using the Microsoft U.S. Building Footprints dataset[29], which contains 129,591,852 computer-generated building polygons derived from satellite imagery via computer-vision methods. Each building polygon is classified by its individual rooftop area into three categories: small (<5,000 ft²), medium (5,000–25,000 ft²), and large (>25,000 ft²). We aggregate results to the ZIP Code level and apply rooftop suitability factors by ZIP Code based on locale type and Census division[30]. We then use nationally derived distributions of roof tilt and azimuth for small, medium, and large buildings to convert suitable rooftop area into panel-placeable area[30]. For flat roofs, panels are assumed to be installed at a tilt equal to latitude, with inter-row spacing sufficient to avoid self-shading at 10:00 a.m. throughout the year[18]. For pitched roofs, panels are assumed to occupy the more south-facing half of the roof, aligned with the roof slope.[31]

## Assessing solar generation potential at the county level

We estimate solar resource availability for each county using hourly meteorological data from the National Solar Radiation Database (NSRDB) for the weather year 2012. The NSRDB provides hourly solar insolation, air temperature, and other relevant meteorological variables at a 9 by 9 km spatial resolution[31]. For utility-scale solar, these data serve as inputs to NREL's System Advisor

Model (SAM)[32], which simulates the hourly output of a single-axis tracking PV system assuming a module efficiency of 15% and computes hourly capacity factors for each grid cell. For rooftop solar, hourly capacity factors are calculated for all azimuth-tilt combinations and then averaged using weights based on NREL's [30] reported distribution of roof orientations for small, medium and large buildings. To obtain county-level values, we calculate the average of all hourly capacity factors within each county, producing a spatially consistent dataset of mean annual capacity factors across the study region.

**Estimating the levelized cost of electricity of solar deployment by county and land use type**

Given electricity generation potential, we calculate the levelized cost of electricity (LCOE) for solar by county and land use type and rooftop size. We calculate the LCOE as:

$$LCOE_{i,l} = \frac{CAPEX_{i,l} \cdot CRF + FOM_{i,l}}{\overline{(E)}_{i,l}}$$

Where $LCOE_{i,l}$ is the levelized cost of electricity (\$/MWh) for county $i$ and land-use type $l$; $CAPEX_{i,l}$ is the capital cost (\$/kW) adjusted for land-use-specific premiums; $CRF$ is the capital recovery factor that annualizes capital costs; $FOM_{i,l}$ is the fixed operation and maintenance cost (\$/MW-yr); and $\overline{E}_{i,l}$ is the average annual electricity generation (MWh) computed from hourly capacity factors.

Capital costs of solar vary between land use types, ground-mounted versus rooftop solar, and locations. With respect to land use types, utility-scale PV development on brownfield sites typically incurs higher capital costs than comparable projects on greenfields. While greenfield projects often benefit from standardized permitting and conventional site preparation and construction techniques, brownfield development involves additional risk mitigation measures (such as remediation of contaminated soils, environmental monitoring, more complex permitting processes, and custom foundation and mounting structures) which translate into higher capital expenditures (CAPEX) [33]. Despite well-documented qualitative cost disparities, publicly available data quantifying the cost differential between brownfield and greenfield solar projects remain limited. Existing datasets [34], [35], assume greenfield development conditions and do not differentiate by land category. To address this data gap, we obtain normalized project-level CAPEX estimates (in $/W-AC) for 56 brownfield projects and 962 greenfield projects between 2010 and 2022 from Lawrence Berkeley National Laboratory (LBNL), which we use to quantify the brownfield cost premium relative to greenfield projects. For each rolling three-year period, we compute the ratio of average CAPEX for brownfield projects to the average CAPEX for greenfield projects. We then take a weighted average across all periods to derive the cost differential. Our analysis reveals that utility-scale PV development on brownfields has, on average, a 30% higher CAPEX compared to greenfields, consistent with values reported in the literature [36]. We apply this 30% premium as a multiplicative factor to the model's greenfield CAPEX inputs from NREL's Annual Technology Baseline (ATB) [34] for development on brownfields, while greenfield project costs remain unadjusted.

Development costs on brownfields are also impacted by project scale, with smaller projects often facing higher costs due to diminished economies of scale [36]. To capture these effects, we draw from empirical trends reported in LBNL's Utility-Scale Solar report [35], which categorize capacity range by their average cost ($/W-DC). We categorize our brownfields into similar capacity bins as defined in the report (5-20 MW, 20-50 MW, 50-100 MW, and 100-700 MW), and then apply the corresponding cost premium to each, thereby accounting for both the brownfield-specific cost adjustment and scale-dependent CAPEX variation. For consistency, we also account for scale-driven cost reductions in greenfield projects using the same capacity bins [35]. Finally, we obtain fixed O&M values for ground-mounted solar from the NREL's ATB [34] , using the "Market" case under a Mid cost scenario (SI, Table S3).

With respect to ground-mounted versus rooftop solar, rooftop solar generally costs more than utility-scale PV because it lacks economies of scale and has a greater share of soft cost [37]. For rooftop PV costs, we use CAPEX and fixed O&M values from the NREL ATB [34]. We assume medium- and large-rooftop systems follow commercial PV costs, while small-rooftop systems follow residential PV costs.

With respect to location, we account for the spatially variable cost of interconnecting the utility-scale facility to the nearest transmission line by applying location specific cost multipliers from NREL's reV model to the national average levelized cost of transmission [38]. These multipliers are derived at a 90 m × 90 m spatial resolution and capture key factors such as terrain, right-of-way constraints, distance to the nearest transmission line, and associated spur line and substation costs [39].

Based on these county-level capacity factors and LCOE estimates, we construct supply curves for each land category by region. These supply curves rank sites in order of increasing LCOE, thereby linking the cumulative technical potential of each land-use category to its associated cost of deployment.

**Scenario framework**

To quantify the effect of land use priorities on power system decarbonization costs, we evaluate scenarios that vary which land categories are prioritized for solar PV siting. We specifically analyze a 2035 deployment target of 453 GW of solar PV capacity across the Eastern Interconnection, consistent with a 100% electricity sector decarbonization pathway [40]. The 453 GW target is divided by RTO territory, with 12 GW in ISO-NE, 92 GW in MISO, 21 GW in NYISO, 42 GW in PJM, 97 GW in SPP, and 189 GW in Southeast (See SI Section 1.3).

When meeting these solar deployment targets, we analyze two families of scenarios for siting of solar: (i) unconstrained optimization, in which PV deployment minimizes LCOE across the entire Eastern U.S. given the Eastern U.S. deployment target; and (ii) regionally constrained optimization, in which PV deployment minimizes LCOE within each region given regional deployment targets. Within each family, we examine four land use prioritization schemes: (i) prioritizing only minimum LCOE; (ii) prioritizing contaminated lands, such that siting is restricted to disturbed lands (brownfields, Superfund sites, landfills, abandoned mine lands, and RCRA Corrective Action sites) until exhausted; (iii) prioritizing greenfield lands; and (iv) prioritizing rooftop solar PV. Comparing these variants quantifies the opportunity costs of prioritizing different solar PV deployment strategies.

**RESULTS**

**Supply curves by land type across the Eastern United States**

Figure 1 illustrates the supply curves for solar PV across land use and rooftop categories in the Eastern U.S., disaggregated by greenfield (**Error! Reference source not found.**a), contaminated sites (**Error! Reference source not found.**b) and building rooftops (4.2c). These supply curves capture the cumulative technical potential and levelized cost of electricity (LCOE) for each land type, after applying uniform siting exclusions and adjusting development costs to account for land-type-specific premiums.

Across all greenfield categories, total solar PV potential exceeds 11.6 TW, far surpassing the deployment target of 453 GW. Prime agricultural land accounts for the largest share (5.6 TW), followed by forested (4 TW) and rangeland/grassland (2 TW). Substantial low-cost potential exists on prime agricultural land, with LCOEs ranging from 39-61 $/MWh up to 453 GW. Rangeland/grassland, forested and barren/marginal lands offer higher LCOEs, ranging from 39-72, 40-57 $/MWh and 41-74 $/MWh, respectively.

Contaminated lands offer considerably less technical potential (about 70 GW or 0.6% of greenfield potential). Within this category, RCRA Corrective Action sites account for the largest share of technical potential (25 GW), followed by Superfund sites (22 GW), brownfields (11 GW), abandoned mines (9 GW), and landfills (0.8 GW). Together, these sites represent less than 18% of the 453 GW solar PV deployment target across the Eastern Interconnect. In addition to lower available capacity, contaminated lands face higher costs relative to greenfield sites, with meaningful variation across land types. The lowest LCOEs on contaminated lands occur on brownfields (~49 $/MWh), followed by RCRA sites (~51 $/MWh), Superfund sites (~52$/MWh), and landfills and abandoned mines (~$57/MWh). These lowest-cost contaminated sites are significantly more expensive than greenfields; approximately 6 TW of solar PV potential is available on greenfields before reaching the lowest-cost contaminated sites. The higher costs on contaminated lands stem from additional financial hurdles related to remediation and development risk mitigation, typically smaller project scale, and, to a partial extent, lower solar PV resource quality tied to the sites (refer to dashed curves in Figure 1d).

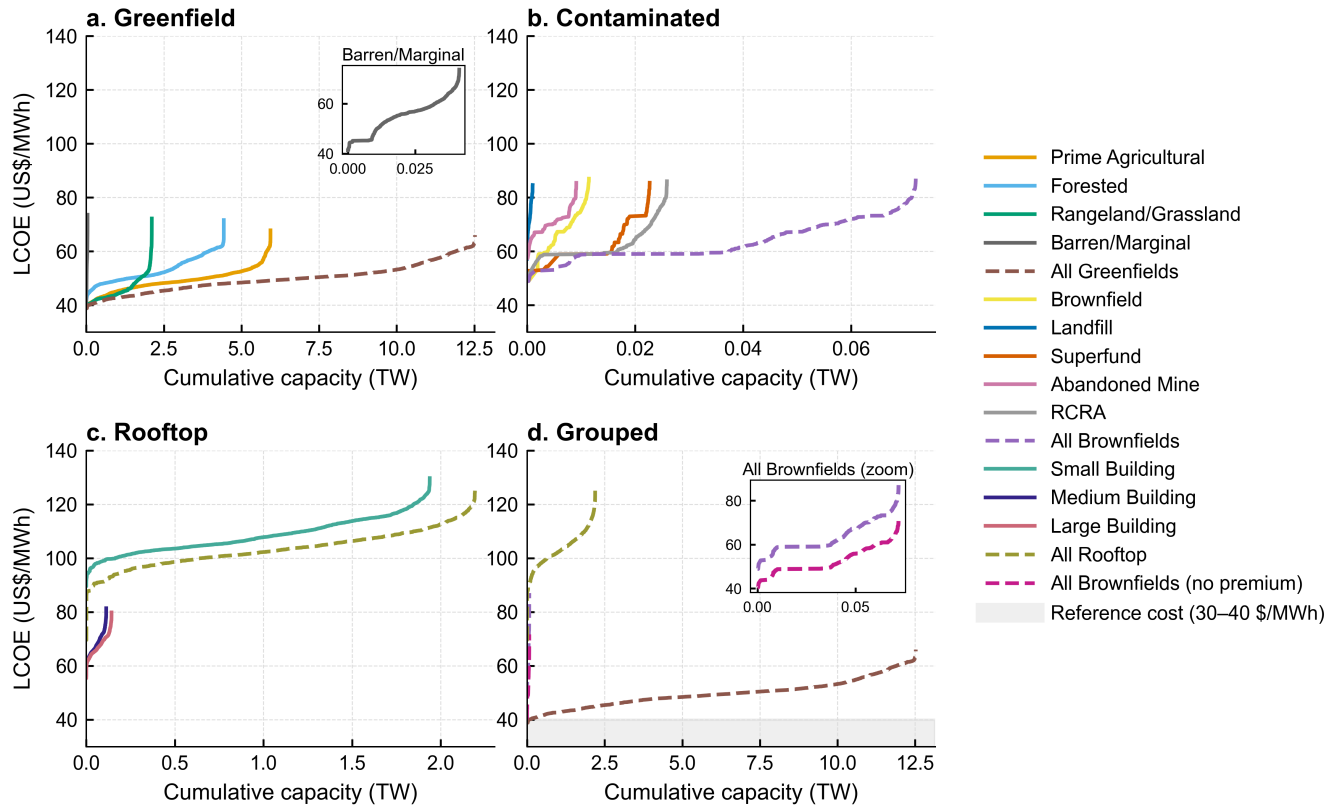

Figure 1: Supply curves for solar PV on land types across the Eastern U.S. The curves show the marginal cost of deploying solar PV across different categories of (a) greenfields, (b) contaminated lands, and (c) rooftops. Panel (d) compares grouped supply curves for all greenfields, brownfields, and rooftops. Note the differences in x-axis scales between panels. Definitions of land-use categories are provided in **Error! Reference source not found.** in the SI.

Rooftop solar offers 2.2 TW of technical potential, larger than that on contaminated lands but smaller than that on greenfields. Among rooftop solar, potential on small rooftops makes up the majority (88%) of total rooftop solar potential, while medium and large rooftops account for 5% and 7%, respectively. However, large and medium rooftops offer lower LCOEs than small rooftops, particularly further along the medium and large rooftop supply curves. From 0 to 453 GW capacity, for instance, the LCOE ranges from 55-80 $/MWh and 56-81 $/MWh for large and medium rooftops, respectively, and $89-116 $/MWh for small rooftops. In comparison, the average LCOEs for all rooftops ($70–108/MWh) are 79–89% higher than those of greenfield sites ($39–57/MWh) for meeting the same 453 GW target, and 24–43% higher than those of contaminated lands ($49–87/MWh) for meeting their total 70 GW capacity.

**Supply curves by land type by regional transmission organization territory**

At the level of regional transmission operator (RTO) territories (Figure 2), both the technical potential and LCOE vary significantly across land use types. To meet regional capacity targets in MISO, SPP, PJM, and the Southeast, prime agricultural land dominates solar PV siting potential, with technical capacities of 3.5 TW in MISO, 3.3 TW in SPP, 1.5 TW in PJM, and 0.7 TW in the Southeast. Large contiguous tracts of cropland in these regions contribute to the lowest LCOEs ($39–48/MWh) for meeting their respective capacity targets (Figure 3, SI Table S3). These regions currently host the majority of utility-scale solar PV development [35], with prime farmland accounting for about 70% of deployments [42]. They also have significant technical potential on forested lands (1–1.7 TW), grasslands (87 GW–1.8 TW), and rooftops (470-920 GW) (Figure 2). The former two land types offer LCOEs comparable to prime agricultural land, while rooftop solar is consistently higher in cost at 83-99 $/MWh to meet each region's capacity target. By contrast, contaminated sites provide limited deployment potential of 21 to 24 GW per region (Table S4 in SI).

In contrast, solar PV potential on prime agricultural land is constrained in ISO-NE and NYISO when compared to MISO, with available capacity dropping by up to 99% (Table S4 in SI). The cost of developing on farmland in ISO-NE and NYISO is also significantly higher than in other regions, reaching $61/MWh and $58/MWh, respectively for capacity targets of 12 GW in ISO-NE and 21 GW in NYISO (Figure 2). However, both NYISO and ISO-NE have just sufficient prime agricultural land to meet their target (Figure 3, Table S4). Higher LCOEs in these regions is driven by lower capacity factors. Both ISO-NE and NYISO also have limited potential on contaminated sites, with only 1.6 GW and 5 GW available, respectively, but much higher potential on rooftops (65 GW in ISO-NE and 90 in NYISO). By comparison, forested lands provide the largest deployment potential, with 224 GW in NYISO and 384 GW in ISO-NE. Costs on forested lands are also relatively lower, with LCOEs 19-35% lower in NYISO and 21-38% lower in ISO-NE compared to brownfields, and 19–48% lower in NYISO and 46–51% lower in ISO-NE compared to rooftops.

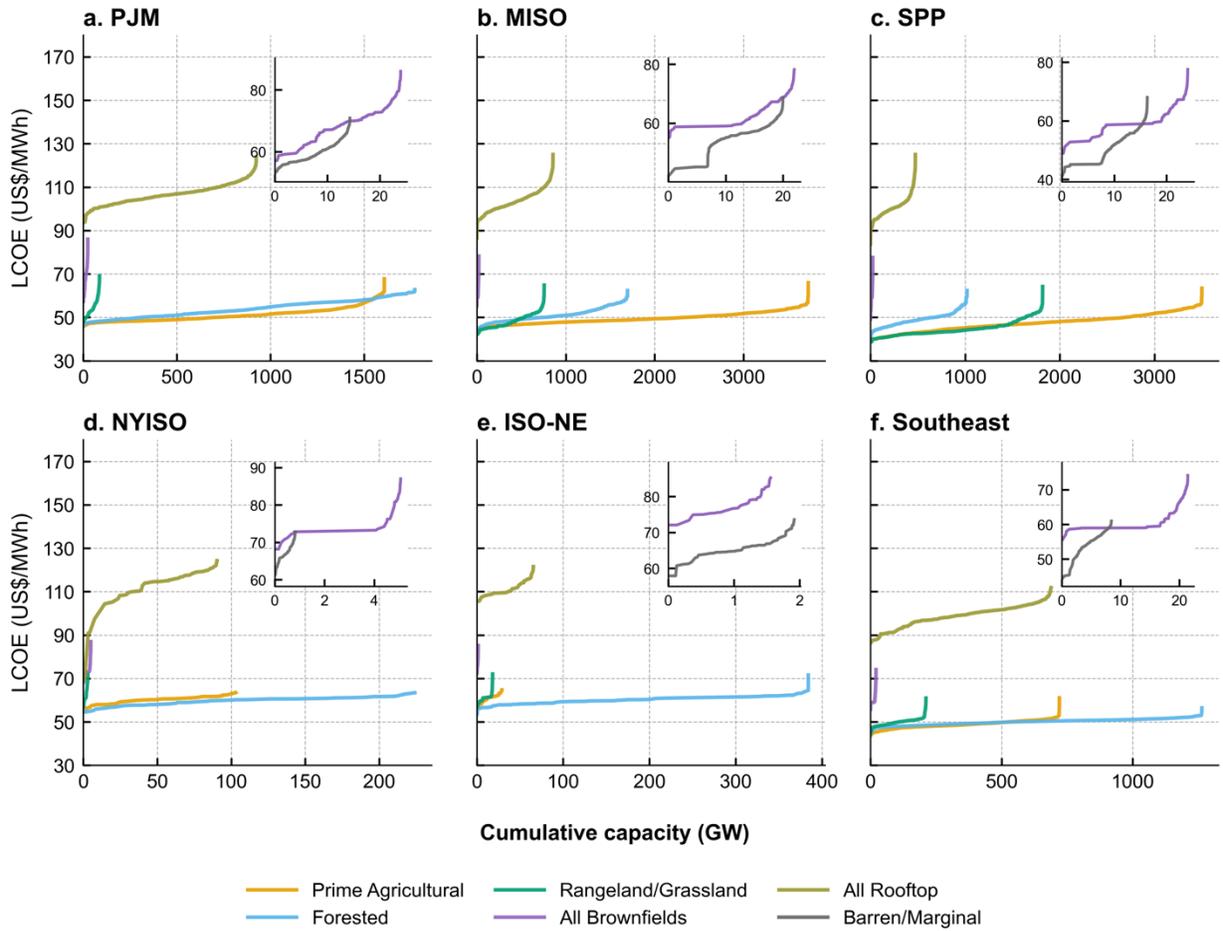

Figure 2. Regional supply curves for utility-scale solar PV by land type across major U.S. power markets. Supply curves show levelized cost of electricity (LCOE) versus cumulative capacity for prime agricultural land, forested areas, rangeland/grassland lands across six regional transmission operator (RTO) territories: PJM, MISO, SPP, NYISO, ISO-NE, and the Southeast. Insets are brownfield and barren/marginal land supply curves separately for visibility. Note differences in x-axis scales between subplots and insets.

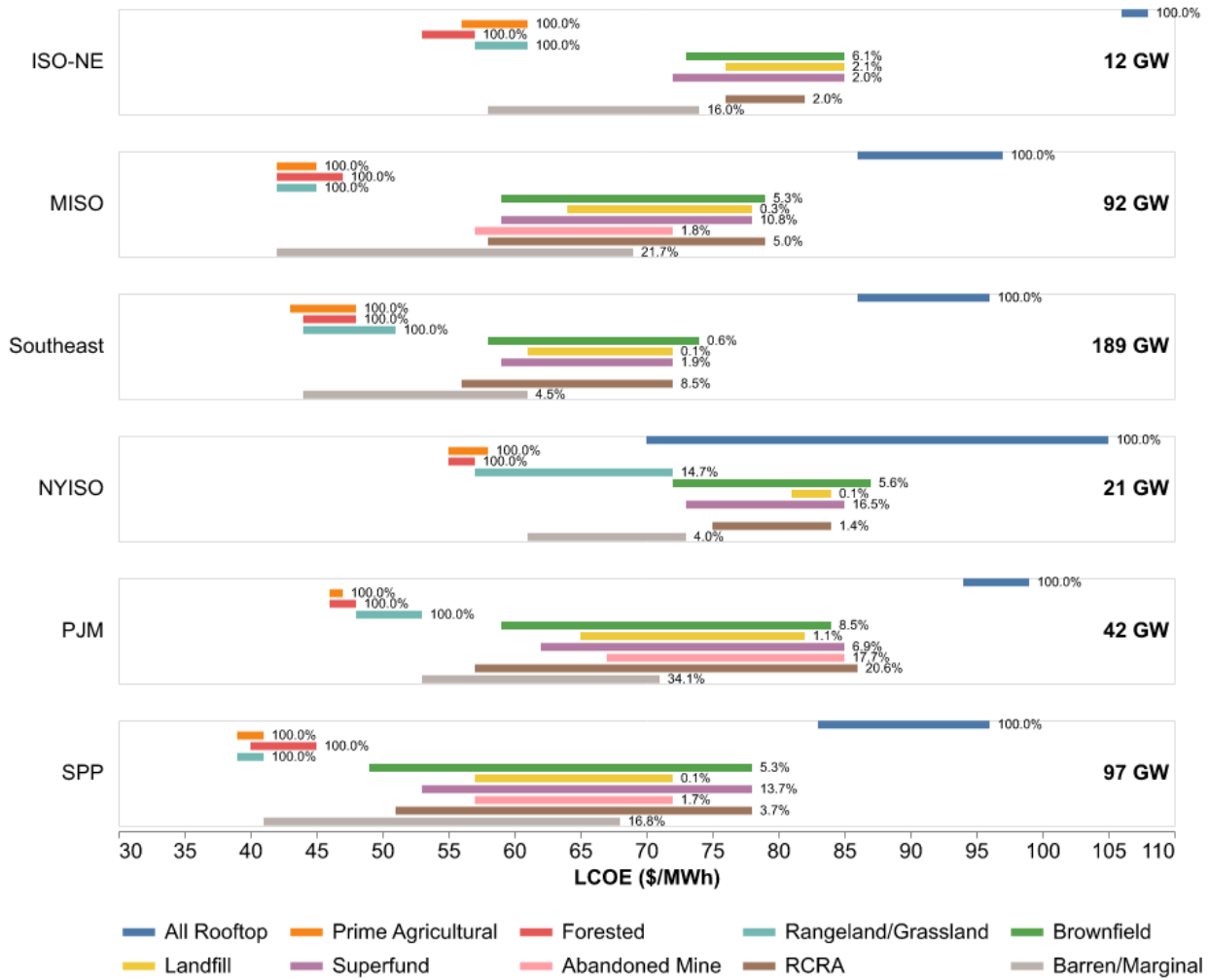

Figure 3. Levelized cost of electricity (LCOE) ranges for solar PV deployment by siting type across regional transmission organizations (RTOs) in the Eastern Interconnection. Colored bars show the minimum–maximum LCOE for each siting category up to each RTO's solar capacity target (indicated by labels on the right). Percentages denote each siting categories' maximum technical potential as a share of each RTO's solar capacity target.

**Costs of alternative land type prioritization schemes for solar PV deployment**

Using our solar PV supply curves, we estimate the costs for achieving a solar PV deployment target of 453 GW under different land type prioritization scenarios (Figure 4). To achieve this deployment level, LCOEs range from $45-112/MWh with a $40-77 billion investment cost across land type prioritization scenarios. Among these scenarios, costs are lowest when prioritizing greenfields, with the average LCOE ranging from 45-47 $/MWh across RTOs. Costs are highest when rooftop development is prioritized, with the average LCOE ranging from 79-101 $/MWh across RTOs.

In scenarios that prioritize contaminated lands, there is not enough technical potential on contaminated lands (70 GW) to host all 453 GW of solar. As a result, a contaminated-lands-first scenario requires deployment in a second category after exhaustion of contaminated lands. For the 453 GW target, deploying on rooftops after contaminated lands ("contaminated then rooftop") results in LCOEs ranging from $74-92/MWh, whereas a "contaminated then greenfield" scenario results in LCOEs of $48-50/MWh. This implies that greenfield land use saving from rooftop application comes at a cost premium of at least $24 per MWh (or $17 billion total cost). In both cases, contaminated-lands-first scenarios yield costs between greenfield and rooftop scenarios.

Under optimal or greenfield-first siting, costs cluster in the $40-57/MWh range, with interior regions (SPP, MISO, PJM, and the Southeast) concentrated near $40-47/MWh, while Northeast regions (ISO-NE, NYISO) sit near $56-57/MWh. Prioritizing deployment on rooftops widens these gaps, increasing LCOEs to $68-70/MWh in interior regions and $84-87/MWh in the Northeast. Emphasizing small rooftops produces the steepest increase in LCOEs, pushing Northeast RTO LCOEs to $114/MWh and interior region LCOEs to 99-104 $/MWh. Contaminated-first pathways fall between these extremes and hinge on the second prioritized land type: contaminated then greenfield outcomes span $44-60 /MWh across regions, whereas contaminated then rooftop spans $65-82/MWh, with the largest penalties concentrated in ISO-NE and NYISO where brownfield capacity saturates early.

In the above results, solar PV deployment is constrained within each RTO to achieve an RTO-based deployment target. If there are no constraints on solar PV deployment spatially, lower LCOE and investment cost are anticipated because new capacity could then shift to concentrated in resource-abundant regions. In this case, more than 90 percent of solar PV will be deployed in the SPP and the Southeast. The overall least-cost LCOE decreases from $45 per MWh to $40 per MWh, which reduces the total capacity cost by 0.37 billion USD. The greenfield, contaminated-then-greenfield, contaminated-then-rooftop, and rooftop-priority scenarios decrease their LCOEs by $4/MWh, $4/MWh, $2/MWh, and $2/MWh, respectively. Thus, the additional cost of preventing spatial concentration of solar deployment is substantially smaller than the additional cost of avoiding greenfield deployment.

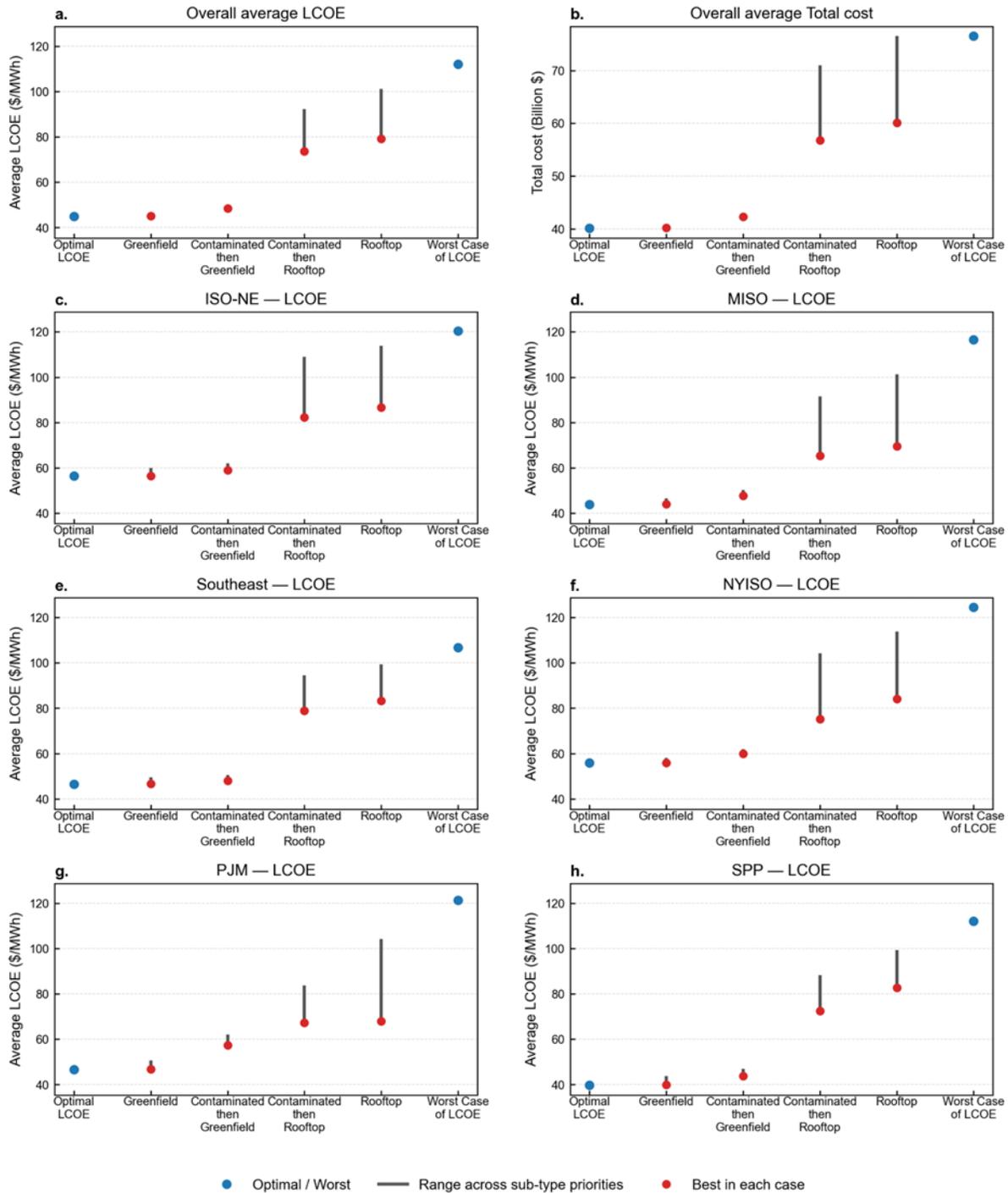

Figure 4: Cost of solar PV deployment towards a 453 GW capacity target across different scenarios. (a&b) show average LCOE and total cost across RTOs, while (c-h) show average LCOE by RTO region. In each plot, blue points indicate the least and highest solar costs across land types; red points indicate the least cost in each deployment scenario; and the grey lines show variability in cost given varying prioritization within land types (e.g., by prioritizing varying rooftop sizes in rooftop scenarios, or greenfield types in greenfield scenarios).

## DISCUSSION

The study quantifies how land use priorities can shape solar PV deployment and costs across the Eastern U.S. We construct and integrate solar PV supply curves by greenfield, brownfield, and rooftop land types. We then quantify solar costs under scenarios that prioritize solar PV deployment on varying land types [43]. Three key messages emerge. First, while contaminated lands are often prioritized in policy for their potential to reduce land-use conflict [44], their technical potential is severely limited. They represent roughly 15% of the 453 GW 2035 regional deployment target, and remain costlier than the least-cost greenfields, with LCOEs of $49–87/MWh compared to $39–57/MWh for greenfields (14-33% less than contaminated sites). Given limited deployment potential, a contaminated-first strategy will not preclude solar PV deployment in other land types, and place the system on a higher cost trajectory than a greenfield-first approach. Second, greenfield sites offer the lowest cost opportunity for meeting capacity targets. Across all categories, greenfields offer more than 12.5 TW of technical potential and include the largest share of sites with costs below $50/MWh. Third, rooftop solar PV, as a land-saving strategy, provides a larger technical potential than contaminated lands (roughly 2.2 TW across the EI) but remains substantially smaller than greenfield sites (<18%), and comes at consistently higher costs, with minimum LCOEs of roughly 70 $/MWh, well above the highest-cost greenfield sites.

These trade-offs are not uniform across RTO regions. In the interior regions (MISO, SPP, PJM and Southeast) abundant cropland and low-impact greenfield sites enable 2035 capacity targets to be met at the lower end of the modeled LCOEs ($39–47/MWh). Avoiding greenfields in these regions would either make it impossible to meet their deployment targets or require costly substitution onto rooftops at $83–99/MWh, adding a $44–52/MWh premium relative to greenfield-focused deployment. In the Northeast (ISO-NE and NYISO), scarce cropland and limited contaminated sites shift least-cost development towards forested sites with mid-range LCOEs. Rooftop potential is higher than these other land types but also more expensive. Consequently, the premium for rooftop-heavy pathways is most acute in ISO-NE and NYISO, whereas in MISO and SPP the premium arises primarily from social or policy constraints rather than land scarcity.

Our analysis suggests a diversified deployment strategy for solar PV could accelerate decarbonization. Such a strategy should treat contaminated sites as a complementary deployment target, maximizing their use where feasible while also developing low impact greenfields. Low impact greenfields could include rangeland and low-productivity grassland, which offer substantial capacity with relatively low opportunity costs compared to cropland or forested lands. By selectively incorporating lower impact greenfield types, states can achieve decarbonization targets at lower system costs while minimizing social and ecological impacts. Moreover, many low impact greenfields are located in rural areas, presenting opportunities for targeted rural development, particularly if solar PV projects are structured to maximize local economic value, such as tax benefits, lease payments and government revenue [45]. Meanwhile, rooftop solar could be considered as a higher cost option under more restrictive siting conditions, for example, if policies strictly limit greenfield conversion or if local opposition rules out other land categories. The selective use of agrivoltaics or managed grazing can further reduce opportunity costs and improve local acceptance, narrowing the gap between inclusive and strict-avoidance strategies.

Expanding transmission offers one of the most effective ways to lower overall system costs and align future deployment with current market prices. Allowing new solar capacity to shift toward solar-resource-rich interior regions lowers systemwide LCOE by about $2–4/MWh, from $45/MWh under regional siting limits to $40/MWh when siting is unconstrained. This corresponds to roughly $0.37 billion in systemwide savings. Greater flexibility also enables access to the lowest-cost greenfield sites, many of which already fall within the 2023 median utility-scale solar LCOE range of $30–40/MWh [41]. However, concentrating deployment in interior regions would result in inequitable hosting and land conversion for PV between regions, raising equity concerns.

Despite the breadth of our geospatial and techno-economic assessment, several factors constrain the scope of our findings. First, the analysis does not incorporate county or township-level zoning ordinances. Local zoning can prohibit utility-scale solar PV on prime farmland or imposing stringent setbacks [46]. Its omission likely overstates the practical capacity available in certain counties on greenfields and brownfields. Second, the RE-Powering dataset, from which we obtain contaminated sites, provides contaminated sites as centroid points, requiring us to create circular buffers to approximate developable area. This approach introduces potential overlap neighboring sites that fall outside contaminated sites, leading to modest uncertainties in capacity estimates on contaminated sites. Third, the study does not embed co-siting strategies into our supply curves that could impact the marginal cost of solar PV development on greenfields. Agrivoltaics or solar PV grazing can increase the premiums for development, but can also lower opportunity costs and enhance local acceptance [47]. Lastly, we treat land categories as fixed through the 2035 planning horizon, even though remediation efforts could gradually reduce the size of developable contaminated sites. At the same time, ecosystem restoration policies might reduce the future availability of certain greenfield types. Future work embedding land use changes over time would refine our results.

## ACKNOWLEDGEMENTS

We thank the Graham Sustainability Institute at the University of Michigan for funding. We are grateful to Sarah Mills for feedback on our analysis; and Joachim Seel and Ben Hoen at Lawrence Berkeley National Laboratory for data on costs of solar PV on brownfields.